\documentclass[12pt,rqno,oneside]{extarticle}

\usepackage{verbatim} 

\usepackage{mathptmx} 

\usepackage{relsize}

\usepackage{titlesec}
\titleformat*{\section}{\Large\bfseries} 
\titleformat*{\subsection}{\large\bfseries} 

\newcommand{\wD}{\widehat D}

\newcommand{\fm}{\mathcal{M}}

\usepackage{cite}
\usepackage[small]{caption}
\usepackage{amsmath}
\usepackage{amssymb}
\usepackage{amsfonts}
\usepackage{setspace}
\usepackage{latexsym}
\usepackage{mathrsfs}
\usepackage{epsfig}
\usepackage{amsthm}
\usepackage{yfonts}
\usepackage{graphicx}
\usepackage{color}
\usepackage{MnSymbol}
\usepackage{psfrag}
\usepackage{enumitem}
\setitemize{wide}

\numberwithin{equation}{section}

\newcommand{\n}{\noindent}
\newcommand{\be}{\begin{equation}}
\newcommand{\ee}{\end{equation}}
\newcommand{\ben}{\begin{displaymath}}
\newcommand{\een}{\end{displaymath}}
\newcommand{\vs}{\vspace{0.2cm}}
\newcommand{\ds}{\displaystyle}

\usepackage{amsbsy}

\usepackage[margin=2cm]{caption} 

\usepackage{fancyhdr}
\AtBeginDocument{\thispagestyle{plain}}
\fancypagestyle{plain}
	{\fancyhead{}
	\fancyfoot{}
	\fancyhead[LE,RO]{}
	\fancyhead[LO,RE]{}
	\fancyfoot[R]{\thepage}
	
	\headsep = 30pt}

\newtheorem{Theorem}{Theorem}[section]

\newtheorem{Definition}[Theorem]{Definition}
\newtheorem{Proposition}[Theorem]{Proposition}
\newtheorem{Conjecture}[Theorem]{Conjecture}

\newtheorem{Aux-Lemma}[Theorem]{Aux-Lemma}

\newcommand{\eexp}{\mbox{\large\ {\rm e}}}
\newcommand{\olfm}{\overline{\mathcal{M}}}
\newcommand{\fv}{\mathcal{V}}

\begin{document}

\n {\huge The area-angular momentum inequality for black holes in cosmological spacetimes} 
\vs\vs

\n {\sc Mar\'ia Eugenia Gabach Clement}, 

\n {\tt gabach@famaf.unc.edu.ar} 
\vs\vs

\n {\sc Mart\'in Reiris}, 

\n {\tt martin@aei.mpg.de}
\vs\vs

\n {\sc Walter Simon}, 

\n {\tt walter.simon@univie.ac.at}

\vspace{0.5cm}

\begin{minipage}[l]{11cm}
\begin{spacing}{.9}{\small For a stable marginally outer trapped surface (MOTS) in an axially symmetric spacetime 
with cosmological constant $\Lambda > 0$ and with matter satisfying the dominant energy condition, we prove that the
area $A$ and the angular momentum $J$ satisfy the inequality  
$8\pi |J| \le A\sqrt{(1-\Lambda A/4\pi)(1-\Lambda A/12\pi)}$
which is saturated precisely for the extreme Kerr-deSitter family of metrics. 
This result entails a universal upper bound $|J| \le J_{\max} \approx 0.17/\Lambda$  
for such MOTS, which is saturated for one particular extreme configuration.  Our result sharpens the inequality  
$8\pi |J| \le A$, \cite{2011PhRvL.107e1101D, 2011PhRvD..84l1503J}, and we follow the overall strategy of its proof in the sense that we first 
estimate the area from below in terms of the energy corresponding to a ``mass functional'', 
which is basically a suitably regularised  harmonic map $\mathbb{S}^2 \rightarrow
\mathbb{H}^2 $. However, in the cosmological case this mass functional acquires an additional
 potential term which itself depends on the area. 
To estimate the corresponding energy in terms of the angular momentum and the cosmological constant 
we use a subtle scaling argument, a generalised ``Carter-identity'',
and various techniques from variational calculus, including the mountain pass theorem.}   
\end{spacing}
\end{minipage}   

\n {\sc PACS}:
02.30.Xx,    
04.70.Bw,
02.40.Hw,
02.40.Ma,
02.40.Vh

\n Keywords: area inequality, apparent horizon, cosmological constant

\section{Introduction}

Some remarkable area inequalities for stable marginally outer trapped surfaces (MOTS) have been proven recently \cite{2011PhRvL.107e1101D}, 
\cite{Dain:2011kb}, \cite{2011PhRvD..84l1503J}, \cite{Simon:2011zf}, \cite{Clement:2012vb}, \cite{Fajman:2013ffa}. 
In particular, for axially symmetric configurations with area $A$ and angular momentum $J$, there is the bound \cite{2011PhRvL.107e1101D}, 
\cite{2011PhRvD..84l1503J}

\be\label{AJ}
|J|\leq \frac{A}{8\pi},
\ee
which is saturated for extreme Kerr black holes. Although a cosmological constant $\Lambda$ does not
explicitly enter into
\eqref{AJ}, this inequality holds in the presence of a non-negative $\Lambda$.
On the other hand, when $\Lambda > 0 $, 
stable MOTS obey the lower bound 
\be\label{ALambda}
A \le 4\pi \Lambda^{-1},
\ee
saturated for the extreme Schwarzschild-deSitter horizon \cite{Hayward:1993tt}. This readily implies the universal
upper bound 
\be\label{JLambda}
|J| \le (2\Lambda)^{-1}
\ee
which, however, can never be saturated even in theory (leaving practical considerations aside in view of the fact that $\Lambda^{-1}$ is of order $10^{122}$.     

The situation bears some analogy to stable MOTS in (not necessarily axially symmetric) spacetimes with 
electromagnetic fields and electric and magnetic charges $Q_E$ and $Q_M$.  
In this case the inequalities $A \ge 4\pi Q^2$ \cite{Dain:2011kb} with $Q^2 = Q_E^2 + Q_M^2$ 
(saturated for extreme Reissner-Nordstr\"om horizons) and $A \le 4\pi
\Lambda^{-1}$ imply the (unsaturated) bound $Q^2 \le \Lambda^{-1}$.
There is however the stronger bound \cite{Simon:2011zf}
\be\label{AQLambda}
\Lambda A^2 - 4\pi A + 16 \pi^2 Q^2 \le 0
\ee
which is saturated for extreme Reissner-Nordstr\"om-deSitter configurations and,
moreover, improves the universal charge bound to $Q^2 \le (4\Lambda)^{-1}$. 

Returning to the present axially symmetric case, the main objective of this article is to incorporate
explicitly the cosmological constant into inequality \eqref{AJ} and determine how it controls the allowed 
values of the angular momentum. We prove the following theorem.
\begin{Theorem} 
\label{main}
Let  ${\cal S}$ be an axially symmetric, stable MOTS together with an
axially symmetric  4-neighborhood of ${\cal S}$ called $({\cal N}, g_{ij})$.
 On $({\cal N}, g_{ij})$ we require Einstein's equations to hold,
with $\Lambda > 0$ and with matter satisfying the dominant energy condition.
Then the angular momentum $J$ and the area $A$ of ${\cal S}$ satisfy 
 \begin{eqnarray}
|J| & \le & \frac{A}{8\pi} \sqrt{\left(1 - \frac{\Lambda A}{4\pi} \right) \left(1 - \frac{\Lambda A}{12\pi} \right)}
\label{A1}\ ,
 \\
|J| &\le & J_{max} = \frac{3 \sqrt{2}}{8 \Lambda \sqrt[4]{3}} \left(1 - \frac{1}{\sqrt{3}} \right) \approx
\frac{0.17}{\Lambda}.
\label{A2}
\end{eqnarray}
Here (\ref{A1}) is saturated precisely for the 1-parameter family of extreme Kerr-deSitter (KdS)  horizons
while  the universal bound (\ref{A2}) is saturated for one particular such configuration.
\end{Theorem}

The proof of this theorem will be sketched in Sect. 
\ref{sec:struc}, while details are postponed to Sect. 5.  
We discuss now its scope and the main differences, similarities and difficulties 
compared  to the ones cited above.

As $\Lambda > 0$, the main inequality \eqref{A1}  is stronger than both 
\eqref{AJ} and \eqref{ALambda}; in particular it forbids the black hole to
rotate as fast as its non-cosmological counterpart.
Concerning the saturation of \eqref{A1}, we observe the same pattern as in the previous inequalities: the extreme
solutions  set a bound to the maximum values of charges and/or angular momentum. 
The non-vanishing cosmological constant does not change this property of extreme black holes.

Inequality \eqref{A2} is obtained in a straightforward manner from \eqref{A1} and makes use of an interesting feature of the extreme KdS family. 
Given $\Lambda>0$ there exists a maximum value for the angular momentum which is attained at a certain value of the area $A$. This property is 
not shared by extreme Kerr horizons ($\Lambda=0$), where the value of $A$ 
determines the angular momentum as $8\pi |J| = A$. Note also that, as opposed to \eqref{JLambda}, \eqref{A2} is sharp and improves the numerical 
factor from 0.5 to 0.17 approximately. 

As stated in Theorem \ref{main}, the inequality \eqref{A1} holds between the area and angular momentum of stable MOTS's. Nevertheless, due to the analogy between stable MOTS 
and stable minimal surfaces in maximal slices, one can prove an analogous 
result for this type of surfaces as well (see \cite{Clement:2012vb} for a discussion of the similarities of these surfaces within the context of geometric inequalities).

Note that matter satisfying the dominant energy condition (DEC) is allowed. The energy condition is required in order to dispose of   the matter terms and to arrive at the 
'clean' inequality \eqref{A1} where matter does not appear explicitly. However, for electromagnetic fields we expect to obtain an inequality between area, 
angular momentum, electromagnetic charges $Q_E$, $Q_M$ and cosmological  constant which should reduce to \eqref{A1} for $Q=0$ and to \eqref{AQLambda} when $J=0$. 
We discuss a corresponding conjecture in Sect. \ref{sec:disc}.  

We now comment on the proof Theorem \ref{main} which is not a straightforward generalisation of previous results.  To
explain this we recall briefly the basic strategy 
of \cite{2011PhRvL.107e1101D}, \cite{2011PhRvD..84l1503J}  that leads to \eqref{AJ}. 
Starting with the stability condition  one obtains a lower bound for the area of the MOTS in terms of a ``mass functional'' $\fm$. This $\fm$ is the key quantity in the proof, 
and  depends only on the twist potential and the norm of the axial Killing vector. The
non-negative cosmological constant and the matter terms
(satisfying the DEC) neither appear in $\fm$ nor later in the discussion in this
case. Therefore, the problem reduces to vacuum and with $\Lambda=0$. Then, a variational principle is used to obtain a lower bound for $\fm$. 
The key point  in this step is the relation between $\fm$ and the ``harmonic energy'' of maps between 
the two-sphere and the hyperbolic plane. This allows to use and adapt a powerful theorem by Hildebrandt et al.
\cite{Hildebrandt}  
on harmonic maps, which gives existence and uniqueness of the minimiser for $\fm$. This minimiser, in turn, gives the right hand side of \eqref{AJ}.

In the present work where we strengthen (\ref{AJ}) to
(\ref{A1}), two important obstacles appear. 
Firstly, the area $A$ now appears not only as upper bound on the corresponding functional $\fm$ but also explicitly in $\fm$ itself. This makes the 
variational principle hard to formulate. We overcome this problem in essence  by ``freezing'' $A$ as well as $J$ to certain values 
corresponding to an extreme KdS configuration, and by adapting the dynamical variables  in $\fm$ suitably.   Secondly, 
the relation of $\fm$ to harmonic maps mentioned above no longer persists, whence the proof of existence and uniqueness of a minimiser for $\fm$ has to be done here from scratch.  
We proceed by proving first that every critical point of $\fm$ is a local minimum. Finally we use the mountain pass theorem in order to get the corresponding global statement.  

Our paper is organised as follows.
 
In Sect. 2 we recall and adapt some preliminary material, in particular the definition of angular momentum 
for general 2-surfaces, as well as the definition of a stable MOTS. 
In Sect. 3 we discuss relevant aspects of the KdS metric, focusing on the extreme case.
In Sect. 4 we sketch the proof of  Theorem \ref{main}, postponing the core of the
argument to three key propositions which are proven in Sect. 5. 

In Sect. 6 we conjecture a generalisation of our inequality to the case with
electromagnetic field along the lines mentioned above already, and we also discuss briefly the case $\Lambda < 0$.

\section{Preliminaries}

\subsection{The geometric setup}

We consider a manifold ${\cal N}$ which is topologically a 4-neighborhood of an embedded 2-surface  ${\cal S}$
of spherical topology.  ${\cal N}$ carries a metric $g_{ij}$ and a Levi-Civita connection $\nabla_i$.  
(Latin indices from $i$ onwards run from 0 to 3, and the metric has signature $(-,+,+,+)$). 
The field equations are
\begin{equation}
\label{ein} 
G_{ij} = - \Lambda g_{ij} + 8\pi T_{ij}
\end{equation} 
where $\Lambda$ is the cosmological constant, and the energy momentum tensor $T_{ij}$ satisfies 
the dominant energy condition. 
In Sections 2 and 3 we allow $\Lambda$ to have either sign; this enables us to
compare with and to carry over useful formulas from work which focuses on
Kerr-anti-deSitter, in particular \cite{Caldarelli:1999xj} and
 \cite{Cho:2008vr}.

We next introduce null vectors  $\ell^i$ and $k^i$ spanning 
the normal plane to ${\cal S}$ and normalized as $\ell^i k_i = -1$.
We denote by $q_{ij} = g_{ij} + 2 l_{(i} k_{j)}$ the induced metric on ${\cal S}$, 
the corresponding Levi-Civita connection by $D_i$ and the Ricci scalar by ${}^2\!R$.
 $\epsilon_{ij}$ and  $dS$ are respectively the volume element and the
area measure on ${\cal S}$. The normalisation $l_i k^i = -1$ leaves a (boost) rescaling freedom $\ell'^i =f \ell^i$, $k'^i = f^{-1}
k^i$. While this rescaling affects some quantities introduced below in an obvious
way,  our key definitions such as the angular momentum (\ref{ang}) and the definition of stability
(\ref{stab}) are invariant, and the same applies to all our results.
The expansion $\theta^{(\ell)}$, the shear $\sigma^{(\ell)}_{ij}$ 
and the normal fundamental form $\Omega_i^{(\ell)}$ 
associated with the null normal $\ell^i$ are given by
\begin{equation}
\label{expsh}
\theta^{(\ell)}=q^{ij}\nabla_i\ell_j \ \ , \ \
\sigma^{(\ell)}_{ij}=  {q^k}_i {q^l}_j \nabla_k \ell_l -
\frac{1}{2}\theta^{(\ell)}q_{ij} \ \, \ \ 
\Omega^{(\ell)}_i = -k^j {q^k}_i \nabla_k \ell_j \ .
\end{equation}

\subsection{Twist and angular momentum}

We now assume that  ${\cal S}$ as well as $\Omega^{(\ell)}_i $ are axially symmetric,
i.e. there is a Killing vector $\eta^i$ on ${\cal S}$ such that 

\begin{equation}
  \label{Lie}
  {\cal L}_\eta q_{ij} = 0 \qquad {\cal L}_\eta \Omega^{(\ell)}_{i} = 0.
\end{equation} 

 The field $\eta^i$ is normalized so that its integral curves have length $2\pi$.

We define the angular momentum of ${\cal S}$ as
\begin{equation}
\label{ang}
J =\frac{1}{8\pi}\int_{\cal S} \Omega_i^{(\ell)} \eta^i dS \ ,
\end{equation}
which will be related to the Komar angular momentum shortly.

By Hodge's theorem, there exist  scalar fields $\omega$ and $\lambda$ on
${\cal S}$, defined up to constants, 
such that  $\Omega^{(\ell)}_{i}$  has the following decomposition 

\begin{equation}
  \label{omlam}
  \Omega^{(\ell)}_{i}=  \frac{1}{2\eta} \epsilon_{ij}D^j \omega +D_j \lambda. 
\end{equation}
From axial symmetry it follows that
\begin{equation}
  \label{eq:16}
 \eta^i \Omega^{(\ell)}_{i}=  
\frac{1}{2\eta} \epsilon_{ij}\eta^i D^j \omega 
 =  \frac{1}{2}\eta^{-1/2}\xi^i D_i  \omega
\end{equation}
where $\eta = \eta^i \eta_i$ and $\xi^i$ is a unit vector tangent to ${\cal S}$ and
orthogonal to $\eta^i$.   

We now recall from \cite{2011PhRvL.107e1101D} that on any axially symmetric 2-surface one can
introduce a coordinate system such that 
 \begin{equation}
\label{can1}
q_{ij}dx^idx^j = e^{2c}e^{-\sigma} d\theta^2 + e^{\sigma} \sin^2 \theta d\varphi^2
 \end{equation}
for some function $\sigma$ and a constant $c$ which is related to the area
 $A$ of ${\cal S}$ via $A = 4\pi e^c$.
In such a coordinate system we can write $J$ as
\begin{equation}
  \label{Jom}
  J =- \frac{1}{8}\int_{0}^\pi  \omega' ~ d\theta
   = -\frac{1}{8} \left[\omega (\pi)-  \omega(0) \right], 
\end{equation}
where here and henceforth a prime denotes the derivative w.r.t. $\theta$.
From now onwards we assume that the Killing vector $\eta^i$
on ${\cal S}$ extends to ${\cal N}$ as a Killing vector of $g_{ij}$. 
Of course this implies (\ref{Lie}). Moreover, it follows that ${\cal L}_{\eta} l = {\cal L}_{\eta} k = 0$.
Using the first equation we obtain   
\begin{equation}\label{etaOmega}
   \eta^i \Omega^{(\ell)}_{i} =  - k^j \ell^i \nabla_i \eta_j. 
\end{equation}
Inserting \eqref{etaOmega} in (\ref{ang}) we see that it indeed coincides with the Komar angular
momentum 
\begin{equation}
\label{AK}
J = \frac{1}{8\pi} \int_{\cal S} \nabla^i \eta^j dS_{ij}.
\end{equation}

We finally introduce the twist vector
\begin{equation}
\omega_i = \epsilon_{ijkl}\eta^{j}\nabla^{k}\eta^l.
\end{equation}
If the energy momentum tensor vanishes on ${\cal N}$, we have  $\nabla_{[i}\omega_{j]} = 0$. 
Hence there exists a twist potential $\omega$, defined up to a constant,
 such that $\omega_i =  \nabla_i \omega$. 
The restriction of this scalar field to ${\cal S}$ is easily seen to coincide with the
 $\omega$ introduced in (\ref{omlam}), which justifies the notation.

In what follows we will refer to the pair $(\sigma,\omega)$ on ${\cal S}$ as
the \textit{data}.

\subsection{Stable marginally outer trapped surfaces} 

We now take ${\cal S}$ to be a marginally trapped surface defined by $\theta^{(\ell)}=0$. 
We will refer to $\ell^i$ as the {\em outgoing} null vector, which leads to
the name marginally outer trapped surface (MOTS).  

Moreover, following \cite{Andersson:2007fh} (Sect. 5) we 
now consider a family of two-surfaces in a neighborhood of ${\cal S}$ 
together with respective null normals $l_i$ and $k_i$ and we impose the following additional requirements on ${\cal S}$
and its neighborhood.

\begin{Definition}
\label{MOTS}
 A marginally trapped surface ${\cal S}$
is stable if there exists an outgoing ($-k^i$-oriented) vector
$X^i= \gamma \ell^i - \psi k^i$, with 
$\gamma \geq 0$ and $\psi > 0$, such that the variation  $\delta_X$
of $\theta^{(\ell)}$ with respect to $X^i$ fulfills the condition

\begin{equation}
\label{stab}
\delta_X \theta^{(\ell)} \geq 0.
\end{equation}

\end{Definition}
 
Two remarks are in order here.
Firstly, it is easy to see (cf. Sect. 5 of \cite{Andersson:2007fh}) that stability of ${\cal S}$
 w.r.t. some direction $X^i$ implies stability w.r.t all directions 
``tilted away from'' $\ell^i$.
In particular, since 
$\delta_{-\psi k} \theta^{(\ell)}    \ge  \delta_X \theta^{(\ell)}$
stability w.r.t. any $X^i$ implies stability in the past
outgoing null direction $-k^i$. 
This latter condition suffices as requirement for all our results.

The other remark concerns the relation between stability and axial symmetry.
We recall that in  \cite{2011PhRvL.107e1101D}, \cite{2011PhRvD..84l1503J}, inequality (\ref{AJ}) was proven under 
the symmetry requirements (\ref{Lie}) and under a stability condition
similar to Definition \ref{MOTS} which, however, required $\psi$ to be
axially symmetric as well. (Axial symmetry of $\gamma$ was also assumed but
not used in the proof). In contrast, in the present theorem (\ref{main}) we
impose the stronger symmetry requirement that  ${\cal S}$
as well as its neighborhood ${\cal N}$ are axially symmetric.
In this case it suffices to impose the stability condition (\ref{MOTS})
as above, namely without explicitly requiring axial symmetry of $\psi$,    
since the existence of an axially symmetric function $\widetilde \psi$
then follows automatically, cf. Thm. 8.2. of  \cite{Andersson:2007fh}.  
Moreover, for {\it strictly} stable MOTS (which satisfy $\delta_X \theta^{(\ell)} \not\equiv 0$
in addition to (\ref{stab})) there follows even axial symmetry of the
surface itself if its neighborhood is axially symmetric (cf. Thm. 8.1. of
\cite{Andersson:2007fh}).

 \section{Kerr-deSitter}
 
 In this section we review some relevant properties of the event horizons of the Kerr-deSitter (KdS) 
solutions, making use of \cite{Caldarelli:1999xj}, \cite{Cho:2008vr}, and references therein.
Other aspects of the rich and complex structure of these spacetimes can be found in
\cite{MR0424186}.
\vs

\subsection{The metric, the horizon and the angular momentum}

 In ``Boyer-Lindquist'' coordinates, the KdS metric is 

\begin{equation}
\label{Kerr}
ds^2 = - \frac{\zeta}{\rho^2} \left(dt - \frac{a \sin^2 \theta}{\kappa} d\phi \right)^2 
+ \frac{\rho^2}{\zeta} dr^2 + \frac{\rho^2}{\chi}
d\theta^2 + \frac{\chi \sin^2 \theta}{\rho^2}  
 \left(a dt - \frac{r^2 + a^2}{\kappa} d\phi \right)^2 
\end{equation}
where 
\begin{eqnarray}
\label{zetrho}
 \zeta & = & (r^2 + a^2)(1 - \frac{\Lambda r^2}{3}) - 2mr,  \qquad \rho^2 = r^2 +
 a^2 \cos^2 \theta \\
\label{kapchi}
\kappa & = & 1 + \frac{\Lambda a^2}{3}, \qquad   \chi = 1 + \frac{\Lambda a^2 \cos^2 \theta}{3}
\end{eqnarray}

 where $m\geq 0$ and $0\leq a^{2}\leq \Lambda^{-1}3(2-\sqrt{3})^{2}$. 

 As a function of $r$, $\zeta$ has one negative root and three positive roots (possibly counted with multiplicities). 
The greatest root, $r_{ch}$, marks the cosmological horizon, while the second greatest, $r_{h}$, 
marks the event horizon (from now on simply called ``horizon'').

The area of the horizon is 
\begin{equation}
\label{AKdS}
A = \frac{4\pi \left(r_h^2 + a^2 \right)}{\kappa}
\end{equation}
and the induced metric on it reads
\begin{equation}
\label{ds}
ds^2 = \underbrace{\frac{\mu_h^2}{\kappa^2 \rho_h^2}}_{e^{\sigma}} 
\bigg(\underbrace{\frac{\kappa^2 \rho_h^4}{\mu_h^2 \chi}}_{e^{2q}} d\theta^2 + \sin^2\theta d\phi^2
\bigg)
\end{equation}
where $\mu_h^2 = (r_h^2 + a^2)^2 \chi$ and $\rho_h =r_{h}^{2}+a^{2}\cos^{2}\theta$. 
 
Hence
\begin{equation}
\label{spq}
e^{\sigma + q} =  \frac{r_H^2 + a^2}{\kappa} = e^c = const. = \frac{A}{4\pi}
\end{equation}   
and the metric is in the "canonical form" (\ref{can1}) of \cite{2011PhRvL.107e1101D} 
\begin{equation}
\label{can2}
ds^2 = e^{\sigma}\left(e^{2q} d\theta^2 + \sin^2\theta d\phi^2 \right)
\end{equation}
 with $\sigma + q = c = const.$.  

We now calculate the twist potential $\omega(\eta)$  everywhere (not
only on ${\cal S}$), for $\eta^a = \partial/d \phi$.
Adapting a known calculation in the case $\Lambda = 0$ (cf. e.g.
Appendix A of \cite{Avila:2008te}   and omitting some intermediate steps,
we find

\begin{eqnarray}
\omega' &=& \omega_{\theta}  =  \epsilon_{\theta\phi r t}g^{rr}g^{tt}\partial_r \eta_t +
  \epsilon_{\theta\phi r t}g^{rr}g^{t\phi}\partial_r \eta_{\phi} =
- \frac{\zeta \sin \theta}{\kappa} \left(g^{tt}\partial_r g_{t\phi} +
 g^{t\phi}\partial_r g_{\phi\phi} \right) = \\
{} & = & - \frac{\kappa}{\chi \sin \theta } 
\left( g_{\phi\phi} \partial_r g_{t\phi} -  g_{t\phi}  \partial_r g_{\phi\phi}
\right) = - \frac{2ma \sin^3 \theta}{\kappa^2 \rho^2} \left[r^2 - a^2
+ \frac{2r^2}{\rho^2}\left(r^2 + a^2 \right) \right] = \label{om1} \\
{} & = & - \frac{2ma}{\kappa^2} \frac{\partial}{\partial \theta} 
\left(\cos^3 \theta - 3 \cos\theta - \frac{a^2 \cos\theta \sin^4 \theta}{\rho^2} \right)  
 \label{om2} 
\end{eqnarray}
It follows that
\be
\label{om3}
 \omega = - \frac{2ma}{\kappa^2}  
\left(\cos^3 \theta - 3 \cos\theta - \frac{a^2 \cos\theta \sin^4 \theta}{\rho^2} \right)  
\ee
We note that compared to the case $\Lambda = 0$,  $\omega$ just gets an extra 
factor $1/\kappa^2$. 
Integrating and using (\ref{Jom}) we obtain in particular that 
\begin{equation}
\label{J}
J = am/{\kappa}^2
\end{equation}
which agrees with Equ. (2.10)  of \cite{Cho:2008vr} and Equ. (18) of \cite{Caldarelli:1999xj}.

\subsection{Extreme horizons}

 When at least two of the three non-negative roots 
of $\zeta(r)$ coincide, (one of which is necessarily $r_{h}$), the horizon is called extremal.
When this happens the geometry near the horizon degenerates to a ``throat''. We refer to \cite{Cho:2008vr} for a further
discussion. In what follows we will just need the relation between the parameters $m,a,\Lambda, A$ and $J$ 
which we derive explicitly.
 
For extremal event horizons the radius of the limiting sphere $r_e$
 satisfies, in addition to  $\zeta (r_e) = 0$, the equation     
\begin{equation}
\label{dp}
0 = \frac{1}{2} \frac{d \zeta}{dr}\bigg|_e = - \frac{2\Lambda r_e^3}{3} + r_e(1 - \frac{\Lambda a^2}{3}) - m.
\end{equation}
Here and henceforth a subscript $e$ indicates extremality.
Eliminating $m$ from $\zeta(r_e) = 0$ and (\ref{dp}) we obtain 
\begin{equation}
\label{re}
\Lambda r_e^4 + r_e^2(\frac{\Lambda a^2}{3} - 1) + a^2 = 0.
\end{equation}
  
For $\Lambda \le 0$ this equation has just a single root which can be
called extremal horizon, while
for $\Lambda > 0$ there are two solutions 
$r_{e} = r_{\pm}$ for given $J$
Explicitly, for  $\Lambda > 0$,  

\begin{equation}
\label{rEpm}
r_{\pm}^2 = \frac{1}{2 \Lambda}\left(1 - \frac{\Lambda a^2}{3} \right) \pm
\frac{1}{2 \Lambda} \sqrt{\left(1 - \frac{\Lambda a^2}{3} \right)^2 - 4 a^2 \Lambda}.
\end{equation}
 When $r_{e}=r_{-}$, (and $r_{e}$ is not a triple root), the first two positive roots meet and $r_{e}<r_{ch}$, which means that a 
 cosmological horizon persists in spacetime. On the other hand when $r_{e}=r_{+}$, then the last two positive root 
coincide and the event and the cosmological horizons become both extremal (and merge). 

Using (\ref{re}) to eliminate $a^2$ from (\ref{AKdS}) we find 
\begin{equation}
\label{ArE}
A = \frac{8\pi r_e^2}{1 + \Lambda r_e^2}.
\end{equation}
On the other hand, eliminating $r_e$ from (\ref{re}) and (\ref{AKdS})  gives 
\begin{equation}
\label{aA}
a^2 = \frac{A}{4\pi} \frac{1 - \Lambda A/4\pi}
{ \left(1 - \Lambda A/8\pi \right) \left(1 - \Lambda A/12 \pi \right)}.
\end{equation}

In equation (\ref{re}) we eliminate now $m$ using (\ref{dp}), then $a^2$
using (\ref{re}) and finally $r_e^2$ using (\ref{re}). We obtain the
following simple relation between the angular momentum and the area for extreme K(a)dS

\begin{equation}
\label{J=A}
|J| = \mathcal{E}(A) := \frac{A}{8\pi} \sqrt{\left(1 - \frac{\Lambda A}{4\pi} \right) \left(1 - \frac{\Lambda
A}{12\pi} \right)}
\end{equation}
which after a trivial reformulation agrees with (2.32) of \cite{Cho:2008vr}. 
In the case $\Lambda > 0$ and $J = 0$ the zeros of the parentheses correspond to the black hole horizon and the cosmological 
horizon of Schwarzschild-deSitter, respectively.

\begin{figure}[h!]
\centering
\begin{psfrags}
\psfrag{J}{$J$}
\psfrag{Jm}{$J_{max}$}
\psfrag{A-}{$A_-(J)$}
\psfrag{A+}{$A_+(J)$}
\psfrag{Am}{$A_{max}$}
\psfrag{A=}{$\frac{4\pi}{\Lambda}$}
\psfrag{J=E}{$J={\cal E}(A)$}
\psfrag{A}{$A$}
\includegraphics[width=10cm,height=7cm]{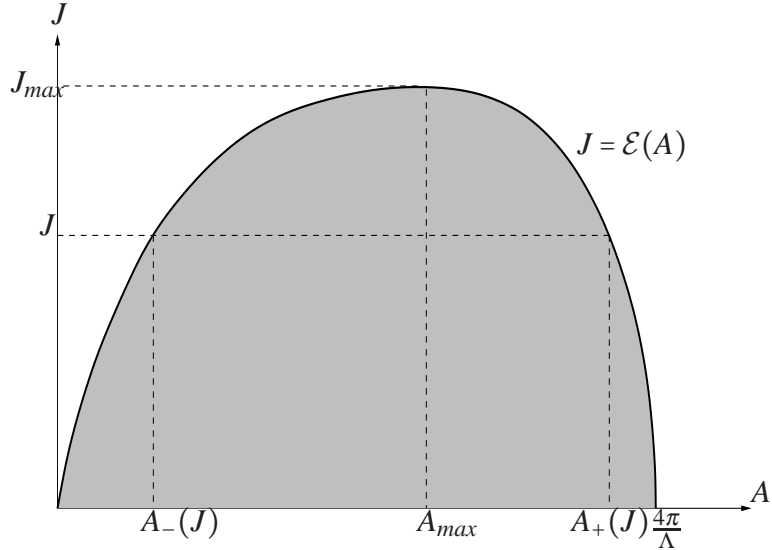}
\end{psfrags}

\caption{The shaded region represents all points satisfying $|J|\leq \mathcal E(A)$.}
\label{fig1}
\end{figure}

For $\Lambda > 0$ we are only interested in the domain 
$\Lambda A/4\pi < 1$ - recall that this bound can be shown for {\it all}~ stable MOTS 
(irrespectively of spherical symmetry) \cite{Hayward:1993tt}. 
In this range of $A$, (\ref{J=A}) takes on a maximal value  
\begin{equation}
J_{max} =  \frac{3 \sqrt{2}}{8 \Lambda \sqrt[4]{3}} \left(1 - \frac{1}{\sqrt{3}} \right) \approx \frac{0.17}{\Lambda}
\quad \mbox{at} \quad  A_{max} = \frac{6\pi}{\Lambda} \left(1 - \frac{1}{\sqrt{3}} \right)
\end{equation}
which is the value stated in (\ref{A2}).
Moreover, for each $J$ with $|J| < J_{max}$ there are {\it two} values $A_-(J) < A_+(J)$ for the
area, cf Fig \ref{fig1}.   

We are now ready to describe the proof of Theorem \ref{main}.

\section{The structure and the proof of the main theorem}\label{sec:struc}

The main inequality
\be\label{MAININ}
|J| \leq  \mathcal{E}(A) 
\ee
with $\mathcal E$ given in \eqref{J=A} and $\Lambda>0$ will not be shown directly but it will follow from a related one.
This is explained in the following Theorem:
\begin{Theorem} 
\label{hat}
For any given MOTS with area $A$, cosmological constant $\Lambda$ and angular momentum $J$,
there is a unique extreme KdS solution with area $\hat A$  constant $\Lambda$ and angular momentum $\hat J$ such that  

\be\label{JJAA}
\frac{|J|}{A^2} = \frac{|\hat J|}{ \hat A^2},
\ee
and  $\hat A \Lambda \le 4\pi$.
Moreover, the inequality $|J| \leq  \mathcal{E}(A)$ is equivalent to the inequality 
\be\label{AAHI}
\hat{A}\geq A.
\ee
\end{Theorem} 

\begin{figure}[h!]
\centering
\begin{psfrags}
\psfrag{J}{$J$}
\psfrag{Jh}{$\hat J$}
\psfrag{Jj}{$J$}
\psfrag{Aa}{$A$}
\psfrag{Ah}{$\hat A$}
\psfrag{J=E}{$J = {\cal E}(A)$}
\psfrag{J=A}{$J = const. A^2$}
\psfrag{A}{$A$}
\includegraphics[width=10cm,height=7cm]{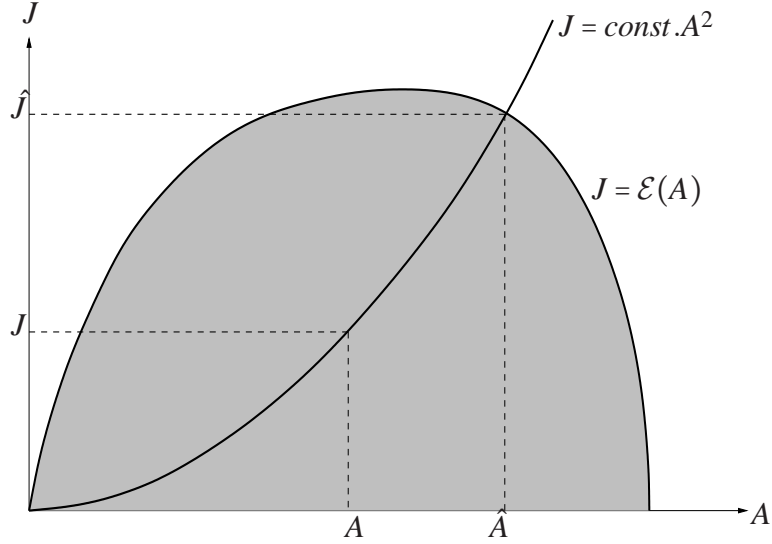}
\end{psfrags}
\caption{The construction described in Theorem \ref{hat}}
\label{fig2}
\end{figure} 

\begin{proof}[\bf Proof] 
The first result, leading to equation \eqref{JJAA},  is intuitively clear from Fig \ref{fig2} since  through any point
$(A,J)$ there is a unique parabola $J/A^2 = const.$, and any such parabola intersects the 
``extreme'' curve $J = {\cal E}(A)$ precisely once apart from the trivial point $(0,0)$.
To state this rigorously, let $\lambda:=A/\hat{A}$ and hence $|\hat{J}|= \lambda^{2} |J|$ and $\hat A \Lambda \le 4\pi$. Then 
the hatted version of (\ref{J=A}) gives a quadratic equation 
 for $\lambda(J,A)$. If  $32 \pi^2 \sqrt{3} |J| >  \Lambda A^2$ this
 equation has a unique solution other than $(0,0)$. Otherwise, there are two
non-trivial solutions but only one of them lies in the region of interest $\hat A \Lambda \le 4\pi$. 

To prove the equivalence between \eqref{MAININ} and \eqref{AAHI}, assume first that $\hat A\geq A$. 
 Then
\be\label{eq0}
A^2\leq \hat A^2=|\hat J|\frac{A^2}{|J|}=\mathcal{E}(\hat A)\frac{A^2}{|J|}=\frac{\mathcal{E}(\lambda A)}{\lambda^2}\frac{\hat A^2}{|J|}
\ee
where we have used \eqref{JJAA}, \eqref{J=A} and $\hat A= \lambda A$,
respectively.  
We next use that the function $\frac{\mathcal{E}(\lambda A)}{\lambda^2}$ is monotonically decreasing 
with $\lambda$ and therefore, as $\hat A\geq A$ we bound the last term as $\frac{\mathcal{E}(\lambda A)}{\lambda^2}\leq \mathcal{E}(A)$.
Putting this together with \eqref{eq0} we find
\be\label{eq2}
\hat A^2\leq \mathcal{E}(A) \frac{\hat A^2}{|J|}
\ee
which gives the desired result, that is, that \eqref{AAHI} implies \eqref{MAININ}.

To prove the converse assume $|J|\leq\mathcal E(A)$. Then $\hat J= \lambda^2 J$ and \eqref{J=A} give 
\be
\mathcal E(\lambda A)= |\hat J|= \lambda^2 |J|\leq \lambda^2\mathcal E(A)
\ee
and therefore
\be
\frac{\mathcal E(\lambda A)}{\lambda^2}\leq \mathcal E(A).
\ee
Again, due to the monotonicity of the left hand side with respect to $\lambda$ we obtain $\lambda A\geq A$ which is \eqref{AAHI}.
\end{proof} 

Having established the equivalence between the main inequality \eqref{MAININ} and \eqref{AAHI}, the next section will be devoted to proving 
\eqref{AAHI} for a stable MOTS ${\cal S}$ with area $A$, angular momentum $J$ and data $(\sigma,\omega)$. 
The proof consists of the same two steps as in the case $\Lambda = 0$. However, as we mentioned in the introduction and 
as we will see below, when $\Lambda>0$ many new complications arise.

In general terms the basic steps can be described as follows.  
\vs

{\rm\bf Step I.} We write the stability inequality  (\ref{stab}) in terms of the data $(\sigma,\omega)$
and multiply it by an axially symmetric function $\alpha^2$ 
whose choice is motivated by the form of the data  $(\sigma,\omega)$ of the extreme KdS horizon.
Then we integrate it on ${\cal S}$ to obtain a lower bound for $A$ 
in terms of the so-called mass functional $\fm$ depending on the dynamic variables $(\sigma,\omega)$. 
 The result is  the following proposition:
\begin{Proposition}\label{PI} Let $(\sigma,\omega)$ be the data of a stable MOTS of area $A$ and angular momentum $J$. 
Then, for any real number $a$ the following inequality holds
\be\label{ineq1}
\frac{A}{4\pi}  \geq   {\rm \eexp}^{\dfrac{{\cal M} (\sigma,\omega, A, a ) - \beta}{\ds 8\kappa}}
\ee
where the functional $\fm$ is given by
\be\label{defM}
\fm(\sigma, \omega, A, a):=\int_{0}^\pi  
\bigg[ \sigma'^2 + 
 \frac{\omega'^2}{\eta^2}
+ 4 \sigma\frac{(1 + \Lambda a^2 \cos^2 \theta)}{\chi} + 4 \bigg(\frac{A}{4\pi}\bigg)^2\Lambda \eexp^{-\sigma} \bigg] \chi\sin\theta d\theta
\ee
 where
 \be
\label{beta}
\beta =  \int_0^{\pi} \left(4 \chi + \frac{\chi'^2}{\chi} \right) \sin \theta d\theta
\ee
and where $\chi(a)$ has been defined in (\ref{kapchi}).

\end{Proposition}

At this stage the constant $a$ is arbitrary, but it will be fixed in the next step.

\vs

{\rm\bf Step II.} 

 The difficulty now is to choose $a$ conveniently and to show that, with such $a$, the r.h.s of (\ref{ineq1}) has the lower 
bound $A^{2}/4\pi \hat{A}$. This would prove (\ref{AAHI}), (hence (\ref{MAININ}) by Theorem \ref{hat}).

We choose $a$ equal to the value that it would take for the extreme black hole of area $\hat{A}$. The explicit form is (\ref{aA}) with
$A$ replaced by $\hat{A}$. We will denote it by $\hat{a}$  
and we denote by $\hat{\kappa}, \hat{\chi}$ and $\hat{\beta}$, the values of $\kappa$, $\chi$ and $\beta$ when $a$ is replaced by 
$\hat{a}$ in (\ref{kapchi}, \ref{beta}). 
Then, for the data $(\sigma, \omega)$ of the given MOTS define   
\be
\label{sigom}
 \hat{\sigma}:=\sigma+2\ln \lambda \qquad
\hat{\omega}= \lambda^{2}\omega,
\ee
where (again) $\lambda = A/\hat{A}$. With this change of variables we obtain 
\be\label{ineq2}
\fm(\sigma,\omega,A,\hat{a})=\fm(\hat\sigma,\hat\omega,\hat{A}, \hat{a})-16\kappa\ln (\hat A/A).
\ee
Thus
\be
\frac{A}{4\pi}\geq \eexp^{\ \dfrac{\fm(\sigma,\omega,A,\hat{a})-\hat\beta}{8\hat\kappa}}=
\left(\frac{A}{\hat{A}}\right)^2\eexp^{\ \dfrac{\fm(\hat\sigma,\hat\omega,\hat{A}, \hat{a})-\hat\beta}{\ds 8\hat\kappa}}
\ee
and we need to prove

\begin{Proposition}\label{PII} In the setup explained above we have 
\be\label{ineq3}
\eexp^{\ \dfrac{\fm(\hat\sigma,\hat\omega,\hat{A},\hat{a})-\hat{\beta}}{8\hat\kappa}}\geq 
\frac{\hat{A}}{4\pi}.
\ee
\end{Proposition}

We wish to mention the following point here. (\ref{ineq3}) means that the
lower bound is obtained by minimising the functional $\fm(\hat\sigma,\hat\omega,\hat{A},\hat{a})$ 
among all pairs $(\hat\sigma,\hat\omega)$ of smooth functions with
$8\pi\hat{J}=-(\hat{\omega}(\pi)-\hat{\omega}(0))$. 
A particular class of such functions has been constructed
above via (\ref{sigom}) from smooth data  $(\sigma,\omega)$ on a smooth MOTS
of area $A$ and angular momentum $J$. However, this does {\it not} mean that 
$(\hat\sigma,\hat\omega)$ will still form smooth data on
a smooth MOTS of area $\hat A$ and angular momentum $\hat J$. This can be seen as follows.      
 In order for the MOTS to be smooth (free of conical singularities), the coordinate function 
$q$ must vanish at the poles, 
i.e. $q(0) = q(\pi)=0$ which implies that $A = 4\pi e^{\sigma(0)} =  4\pi
e^{\sigma(\pi)}$. But inserting the scaling law (\ref{sigom}) in the latter relation contradicts the smoothness property 
$\hat A = 4\pi e^{\hat \sigma(0)} = 4 \pi e^{\hat \sigma(\pi)}$  for the
hatted data, (except in the trivial case $\lambda = 1$).
  Therefore, $\fm(\hat\sigma,\hat\omega,\hat{A},\hat{a})$ should
be considered as 'abstract' functional in the sense that its arguments are no longer
 directly related to any MOTS. Nevertheless, extreme KdS is not only a critical point of
$\fm(\sigma,\omega,A,a)$ but also of $\fm(\hat\sigma,\hat\omega,\hat{A},\hat{a})$, and the properties of the
latter functional enable us to prove (\ref{ineq3}).

\vs

Next we present the proofs of Propositions \ref{PI} and \ref{PII}.

\section{Proof of the main propositions}

\subsection{Proof of Proposition \ref{PI}}

\begin{proof}[\bf Proof.] 
The proof is analogous to the case $\Lambda = 0$ \cite{2011PhRvD..84l1503J} 
to which it reduces by setting $\chi \equiv 1$. The starting point is the stability inequality (\ref{stab})
in which we take $\psi$ to be axially symmetric without loss of generality
(cf. the remarks after Definition \ref{MOTS}).
In terms of  the quantities introduced in Sect. 2 we obtain, integrating
(\ref{stab}) against any axisymmetric function $\alpha:{\mathcal S}\rightarrow \mathbb{R}$,
\be
\label{PEP}
\int_{\cal S} ( |D \alpha|^2 + \frac{{}^2R}{2} \alpha^2 - \alpha^2 |\Omega^{(\ell)}|^{2} - \Lambda \alpha^2 ) dS \geq 0.
\ee
As mentioned in the previous section, we choose the trial function based on the form of
the extreme KdS geometry as
 \be
\alpha=\chi^{1/2}\eexp^{-\sigma/2}.
\ee
In the coordinates (\ref{can2}) the scalar curvature takes the form
\be
{}^{2}R=\frac{\eexp^{\sigma-2c}}{\sin\theta}\bigg[-2\sigma' \cos\theta - \sin\theta \sigma'^{2}+2\sin\theta - 
(\sin\theta \sigma')'\bigg].
\ee
Using this expression we obtain
\begin{align}
\label{PEQ} 
\frac{1}{2\pi} \int_{\cal S} \bigg( |D \alpha|^2 +
\frac{{}^2R}{2} \alpha^2\bigg) dS = & \int_0^{\pi} \bigg(\frac{\chi  \sigma'^2}{4}  - \frac{\sigma' \chi' }{2} + \frac{\chi'^2}{4\chi} \bigg)\sin\theta d\theta \\
\label{SEQ} & \int_0^{\pi} \bigg(-\chi \sigma'  \cos \theta - \frac{\chi \sigma'^2 \sin \theta}{2} + \chi  \sin \theta
 - \frac{\chi  \left(\sin\theta \sigma' \right)'}{2} \bigg) d\theta. 
\end{align}
Integration by parts and some rearrangement yields 
\begin{align}
\label{PEQQ} 
\frac{1}{2\pi} \int_{\cal S} \bigg( |D \alpha|^2 + \frac{{}^2R}{2} \alpha^2\bigg) dS  = & 
-\int_{0}^{\pi}\bigg[\frac{\sigma'^{2}}{4}+\sigma (1 + \frac{2\Lambda a^{2}}{3}\cos\theta)\bigg] \sin\theta d\theta\\
\label{PEQQQ}& + \int_{0}^{\pi}\bigg(\chi+\frac{\chi'^{2}}{4\chi}\bigg)\sin\theta d\theta - \chi\sigma\cos\theta\bigg|_{0}^{\pi}.
\end{align}
Using (\ref{spq}), the last term in line (\ref{PEQQQ}) above is equal to $2 \kappa \ln (A/4\pi)$. Finally, still following \cite{2011PhRvD..84l1503J}, we have   
\be
-\frac{1}{2\pi} \int_{\cal S} (\alpha^2 |\Omega^{(\ell)}|^{2} + \Lambda \alpha^2) dS = -\frac{1}{4} \int_0^{\pi} \chi \frac{\omega'^2}{e^{2\sigma} \sin^4\theta}   
\sin\theta d\theta - \Lambda e^{2c}  \int_0^{\pi} \chi e^{-\sigma} \sin \theta d\theta.
\ee
Combining equations (\ref{PEP}), (\ref{PEQQ})-(\ref{PEQQQ}) and (\ref{PQIV})
we find
\be\label{PQIV}
2\kappa \ln \bigg(\frac{A}{4\pi}\bigg)\geq \frac{\fm - \beta}{4}
\ee
with $\beta$ as in (\ref{beta}). This expression is equivalent to (\ref{ineq1}) as wished. 
\end{proof} 

\subsection{Proof of Proposition \ref{PII}}

In this section we prove \eqref{ineq3} where the hatted variables $(\hat\sigma,\hat\omega)$ refer to the rescaled quantities introduced in \eqref{sigom}. To simplify the notation, for this section only, we omit the hats on these functions. With the new notation, inequality \eqref{ineq3} reads
\be\label{ineq4}
\eexp^{\ \dfrac{\fm(\sigma,\omega,\hat{A},\hat{a})-\hat{\beta}}{8\hat\kappa}}\geq 
\frac{\hat{A}}{4\pi}.
\ee

As in the proof of the inequality in the $\Lambda=0$ case, this step is done by minimising the functional $\mathcal M$. We find first a minimum of $\fm$ for functions $\sigma,\omega$ defined on compact intervals $[\theta_a,\theta_b]\in(0,\pi)$ (in Prop. 
\ref{LOCMIN} and \ref{GLOBAMIN}), and then take the limit 
$[\theta_a,\theta_b]\to[0,\pi]$ to find \eqref{ineq4} (in Prop. \ref{Limit}). Recall that when $\Lambda=0$
the extreme Kerr geometry is the minimiser of the corresponding functional.

In this $\Lambda>0$ case, we find by a straightforward computation that extreme KdS data $(\sigma_e,\omega_e)$ 
is a critical point of $\fm$, that is, the explicit functions

\be
\label{EXTR}
e^{\sigma_e} = \frac{\hat \mu_e^2}{\hat \kappa^2 \hat \rho_e^2},\qquad
\omega_e'=\frac{-2 \hat \chi \hat a \hat r_e(\hat r_e^2+ \hat a^2)^2\sin^3\theta}{\hat \mu_e
\hat \rho_e^4}
\ee
satisfy the Euler-Lagrange equations of $\fm$:
\be\label{EL1a}
\frac{1}{\sin \theta}\frac{d}{d\theta}(2\hat \chi\partial_\theta\sigma\sin\theta)=-\frac{2
\hat \chi\omega'^2}{\eta^2} + 4(1 + \Lambda \hat a^2 \cos^2 \theta)- 
\frac{\Lambda \hat \chi \hat A^2}{4\pi^2}e^{-\sigma}
\ee
\be\label{EL2a}
\frac{d}{d\theta}\left(\sin\theta\frac{\hat \chi \partial_\theta\omega}{\eta^2}\right)=0.
\ee
In (\ref{EXTR}), the quantities $\hat \rho_e$, $\hat \kappa_e$, $\hat \mu_e$
$\hat r_e$ and $\hat \chi_e$ were defined in (\ref{zetrho}), (\ref{kapchi}),
 below (\ref{ds}) and in (\ref{re}) but carrying  subscripts and
 hats they refer here to the extreme KdS solution with parameter $\hat a$. 
Using (\ref{re}) it is easy to
see that the above $\omega_e'$ indeed coincides with (\ref{om1}) and therefore with (\ref{om2}).

This property of extreme KdS geometry will play a fundamental role in the proof of \eqref{ineq4}, but before going into details, some
preliminary definitions are needed.


\vs

{\bf Preliminaries}. Let $0<\theta_{a}<\theta_{b}<\pi$ be fixed. 
For any function $f:[\theta_{a},\theta_{b}]\rightarrow \mathbb{R}$ in $H^{1,2}$ define
\begin{align}\label{NOTAT}
& \|f\|^{2}_{2}:=\|f\|^{2}_{L^{2}}=\int_{\theta_{a}}^{\theta_{b}}f^{2}\ d\theta,\\ 
& \|f\|^{2}_{1,2}:=\|f\|_{H^{1,2}}=\int_{\theta_{a}}^{\theta_{b}} \left[(\partial_\theta f)^{2}+f^{2}\right]\ d\theta=\|\partial_\theta f\|^{2}_{2}+\|f\|_{2}^{2}.
\end{align}
Then, for any $\theta_{1}<\theta_{2}$, ($\theta_{a}<\theta_{1}$ and $\theta_{2}<\theta_{b}$), we have
\be
|f(\theta_{1})-f(\theta_{2})|^{2}\leq |\theta_{2}-\theta_{1}| \|f\|_{1,2}^{2}.
\ee
This says in particular that $f$ is uniformly continuous and we have
\be\label{REFBEL}
\|f-f_{a}\|^{2}_{\infty}:=\sup\big\{ (f-f_{a})^{2}(\theta): \theta\in [\theta_{a},\theta_{b}]\big\}\leq \pi \|\partial_\theta f\|^{2}_{2}\leq \pi \|f\|_{1,2}^{2} 
\ee
where $f_{a}=f(\theta_{a})$.

We will use the affine space $\Gamma_{ab}$ of $H^{1,2}$ paths $\gamma:[\theta_{a},\theta_{b}]\rightarrow \mathbb{R}^{2}$, $\gamma=(\sigma,\omega)$, such that 
\be\label{BOUNDAT}
(\sigma(\theta_{a}),\omega(\theta_{a}))=(\sigma_{e}(\theta_{a}),\omega_{e}(\theta_{a}))\quad \text{and}\quad (\sigma(\theta_{b}),\omega(\theta_{b}))=(\sigma_{e}(\theta_{b}),\omega_{e}(\theta_{b})),
\ee
where $(\sigma_{e},\omega_{e})$  are the data of extreme KdS of area $\hat A$. 

In line with the notation (\ref{NOTAT}) we use the
shorthand  $\| \gamma_{1}-\gamma_{2}\|^{2}_{1,2}:=\|\sigma_{1}-\sigma_{2}\|_{1,2}^{2}+\|\omega_{1}-\omega_{2}\|_{1,2}^{2}$.

Let $\fm_{ab}=\fm_{ab}(\gamma):\Gamma_{ab}\rightarrow \mathbb{R}$ be the functional given by
\be\label{FMAB}
\fm_{ab}(\gamma)=\int_{\theta_{a}}^{\theta_{b}} \bigg((\partial_{\theta}\sigma)^{2}+
4\sigma\frac{(1+\Lambda \hat{a}^{2}\cos^{2}\theta)}{\hat\chi}+
\frac{(\partial_{\theta}\omega)^{2}}{\eexp^{2\sigma}\sin^{4}\theta}+
4\Lambda \bigg(\frac{\hat A}{4\pi}\bigg)^{2}\eexp^{-\sigma}\bigg)\, \hat\chi\, \sin\theta d\theta.
\ee
Note that this functional is the same as the $\fm$ appearing in \eqref{ineq4} except that the integration is over $[\theta_{a},\theta_{b}]$ and that the arguments 
$\gamma=(\sigma,\omega)$ vary in $\Gamma_{ab}$.

\vs
{\bf The functional $\olfm_{ab}$}. Consider the change of variables $(\theta,\sigma,\omega)\rightarrow 
(\bar{\theta}, \bar{\sigma},\bar{\omega})$ given by
\be\label{CHANVAR}
\frac{d\bar{\theta}}{d\theta}=\frac{\sin\bar{\theta}}{\sin\theta\hat\chi(\theta)},\quad \bar{\sigma}=
\sigma+2\ln \frac{\sin\theta}{\sin\bar{\theta}},\quad \bar{\omega}=\omega.
\ee

Explicitly,  $\bar \theta(\theta)$ reads, with a suitable choice of the integration
constant,
\begin{equation}
\label{tr}
\tan \frac{\bar \theta}{2} = 
 \left( \tan \frac{\theta}{2} \right)^{1/\hat \kappa} 
\exp \left[ - \frac{\hat a}{\hat \kappa} \sqrt{\frac{\Lambda}{3}} 
\arctan \left( \hat a  \sqrt{\frac{\Lambda}{3}} \cos \theta \right) \right].    
\end{equation}

It follows that the map $\theta\rightarrow \bar{\theta}$ is a diffeomorphism from $[0,\pi]$ into $[0,\pi]$ and that 
$0<c_{1}<(\sin\theta/\sin\bar{\theta})<c_{2}<\infty$ for $c_{1}$ and $c_{2}$ depending only on $\hat{a}^{2}\Lambda$.  
The transformation of the affine space $\Gamma_{ab}$ will be denoted by $\overline{\Gamma}_{ab}$. A straightforward computation shows 
\begin{eqnarray}
\label{MM}
\fm_{ab}(\gamma)&=&\olfm_{ab}(\bar\gamma)+\\\nonumber
&+&\int_{\bar\theta_{a}}^{\bar\theta_{b}}\frac{4\cos^2\bar\theta}{\sin\bar\theta}d\bar\theta+4\bar\sigma\cos\bar\theta\bigg|_{\bar\theta_{a}}^{\bar\theta_{b}}-
\left.4\sigma(\cos\theta+\frac{\hat{a}^2\Lambda}{3}\cos^3\theta)\right|_{\theta_a}^ {\theta_b}-\int_{\theta_a}^ {\theta_b}\frac{4\hat\chi \cos^2\theta}{\sin\theta} d\theta
\end{eqnarray}
where the functional $\olfm_{ab}=\olfm_{ab}(\bar{\gamma}):\overline{\Gamma}_{ab}\rightarrow \mathbb{R}$ is given by
\be\label{OLFM}
\olfm_{ab}(\bar{\gamma})=\int_{\bar{\theta}_{a}}^{\bar{\theta}_{b}}\bigg((\partial_{\bar{\theta}}\bar{\sigma})^{2}+
4\bar{\sigma}+\frac{(\partial_{\theta}\bar\omega)^{2}}{\eexp^{2\bar\sigma}\sin^{4}\bar\theta}+
\bigg[4\Lambda \bigg(\frac{\hat A}{4\pi}\bigg)^{2}\frac{\sin^{4}\theta}{\sin^{4}\bar\theta}\hat{\chi}^{2}(\theta)\bigg] 
\eexp^{-\bar\sigma}\bigg)\sin\bar\theta d\bar\theta.
\ee
Thus, the functionals $\fm_{ab}:\Gamma_{ab}\rightarrow \mathbb{R}$ and $\olfm_{ab}:\overline{\Gamma}_{ab}:\rightarrow \mathbb{R}$ differ by a constant and boundary terms. 
This immediately implies that $\gamma$ is a critical point of $\fm_{ab}$ iff $\bar{\gamma}$ is a critical point of $\olfm_{ab}$.  In particular as $\gamma_e$ is a critical point of 
$\fm_{ab}$,  $\bar\gamma_e$ is a critical point of $\olfm_{ab}$. As we will explain below, the nature of 
critical points of the functional $\olfm_{ab}$ can be easily analysed via a crucial formula due to Carter. A similar simple formula to analyse the critical points of $\fm_{ab}$ is 
unknown to us. For this reason we will continue working with $\olfm_{ab}$ rather than with $\fm_{ab}$.

\vs
{\bf The results}. The next three propositions together prove Proposition \ref{PII}. 
Propositions \ref{LOCMIN} and \ref{GLOBAMIN} deal with the minimisation of the restricted functional $\olfm_{ab}$. Then,
Proposition ref{Limit} establishes  the connection between 
the minimisation  of $\olfm_{ab}$ (or, equivalently, the minimisation of $\fm_{ab}$) and the minimisation of the original 
functional $\fm$ that ultimately leads to \eqref{ineq4} and Proposition \ref{PII}. The angles $\theta_{a},\theta_{b}\in(0,\pi)$  defining $\olfm_{ab}$ are arbitrary. 

\begin{Proposition}\label{LOCMIN}
For any critical point $\bar{\gamma}_{c}$ of $\olfm_{ab}$ there are constants $\epsilon>0$ and $c>0$, 
such that if $\|\bar\gamma - \bar{\gamma}_{c}\|_{1,2}\leq \epsilon$ then
\be\label{PROPINEQ}
\olfm_{ab}(\bar\gamma)\geq \olfm_{ab}(\bar{\gamma}_{c})+c\|\bar\gamma-\bar{\gamma}_{c}\|^{2}_{1,2}.
\ee
In particular $\olfm_{ab}$ achieves a strict local minimum at any of its critical points.
\end{Proposition}

\begin{Proposition}\label{GLOBAMIN} $\olfm_{ab}$ has only one critical point $\bar{\gamma}_{c}=\bar{\gamma}_{e}$ and 
$\olfm_{ab}(\bar{\gamma}_{e})$ is a global minimum, \textit{i.e.}
\be\label{eqprop5.2}
\olfm_{ab}(\bar\gamma)\geq \olfm_{ab}(\bar{\gamma_e}).
\ee
\end{Proposition}

\begin{Proposition}\label{Limit}
 We have
\be\label{eqprop5.3a}
\fm_{ab}(\gamma,\hat{A},\hat{a})\geq \fm_{ab}(\gamma_e,\hat{A},\hat{a})
\ee
for functions $\gamma=(\sigma,\omega)$ having the boundary values $\gamma|_{\theta_a,\theta_b}=\gamma_e|_{\theta_a,\theta_b}$.

Moreover, taking the limit $[\theta_a,\theta_b]\to[0,\pi]$ we have
\be\label{eqlimit}
\fm(\gamma,\hat{A},\hat{a})\geq \fm(\gamma_e,\hat{A},\hat{a}).
\ee
The explicit form of $\fm(\gamma_e)$ gives \eqref{ineq4}.
\end{Proposition}
Note that taking the limit $(\theta_a,\theta_b)\to (0,\pi)$ is a very delicate issue as the limit boundary values of
$\sigma$ are not necessarily the same as those of $\sigma_e$. We will treat this problem following the ideas of \cite{2011CQGra..28j5014A}.

\begin{proof}[\bf Proof of Proposition \ref{LOCMIN}.] For given $\bar\gamma$ let $\tilde\gamma=(\tilde\sigma,\tilde\omega):=
\bar\gamma-\bar\gamma_{c}$ and define the path $\bar\gamma_{\tau}=\bar{\gamma}_{c}+\tau \tilde\gamma$ for $\tau$ in $[0,1]$. The Taylor expansion of $\olfm_{ab}(\bar\gamma_{\tau})$ at $\tau=0$ gives
\be\label{TAY}
\olfm_{ab}(\bar\gamma)=\olfm_{ab}(\bar{\gamma}_{c})+\frac{1}{2}\partial^{2}_{\tau}\olfm_{ab}(\bar{\gamma}_{\tau})\big|_{\tau=0}+\frac{1}{6}\partial^{3}_{\tau}\olfm_{ab}(\bar\gamma_{\tau})\big|_{\tau=\tau^*}
\ee
where $0\leq \tau^{*}\leq 1$. The proof of Proposition (\ref{LOCMIN}) comes from analysing the last two terms
on the right hand side of (\ref{TAY}). We do that separately.

To simplify notation set $\olfm_{ab}(\bar{\gamma}_{\tau})=\olfm_{ab}$. 
Moreover, in the present proof primes on functions denote derivatives $\partial_{\bar \theta}$.

The first $\tau$-derivative of $\olfm_{ab}$ as a function of $\tau$ is
\be
\partial_{\tau}\olfm_{ab}=2\int_{\bar{\theta}_{a}}^{\bar{\theta}_{b}} 
\left[\wD\tilde\sigma\cdot \wD\bar\sigma+2\bar\sigma+\frac{ \wD\tilde\omega\cdot
\wD\bar\omega-
\tilde\sigma(\wD\bar\omega)^2}{\bar\eta^2}- \frac{\tilde\sigma\fv\eexp^{-\bar\sigma}}{2}\right]\sin\bar\theta d\bar\theta,
\ee
where 
\be
\fv:=4\Lambda \bigg(\frac{\hat A}{4\pi}\bigg)^{2}\frac{\sin^{4}\theta}{\sin^{4}\bar\theta}\hat{\chi}^{2}(\theta)
\ee
and the derivative operator $\wD$ and the dot products are taken with respect to the standard metric on
$S^2$. (Due to axisymmetry $\wD=\partial_{\bar{\theta}}$).
Evaluate at $\tau=0$, integrate by parts and use the boundary conditions to obtain the Euler-Lagrange equations 
for $\olfm_{ab}$, namely
\begin{align}
\label{EL1} & \widehat\Delta\bar\sigma_c-2+\frac{(\wD\bar\omega_c)^2}{\bar\eta_c^2}=-\frac{\fv}{2}\eexp^{-\bar\sigma_c}, \\
\label{EL2} & \wD\bigg(\frac{ \wD\bar\omega_c}{\bar\eta_c^2}\bigg)=0,
\end{align}
where $\bar\eta_c=e^{\bar\sigma_c}\sin^2\bar\theta$, and $\widehat \Delta$ is the Laplace operator with respect to the 
standard metric on $\mathbb{S}^2$. (Again, due to axisymmetry, $\widehat \Delta$ involves only derivatives with respect to $\bar\theta$).

The second $\tau$-derivative of $\olfm_{ab}$ reads
\be
\partial_{\tau}^{2}\olfm_{ab}=2\int_{\tilde\theta_{a}}^{\tilde\theta_{b}}\left[(\wD\tilde\sigma)^2+\frac{2\tilde\sigma^2( 
\wD\bar\omega)^2-4\tilde\sigma \wD\bar\omega\cdot 
\wD\tilde\omega+(\wD\tilde\omega)^2}{\bar\eta^2}+\frac{\tilde\sigma^2\fv \eexp^{-\bar\sigma}}{2}\right]\sin\bar\theta d\bar\theta.
\ee

Next, recall Carter's identity in the form (see \cite{Dain:2010qr})
\be
\label{Carter}
F+ \tilde\sigma G'_{\bar\sigma}+\tilde\omega G'_{\bar\omega}+2\tilde\sigma\tilde\omega G_{\bar\omega}-
{\bar\eta^{-2}}\tilde\omega^2G_{\bar\sigma}=H,
\ee
where

\begin{align}
& G_{\bar\sigma}(\tau)=\widehat \Delta\bar\sigma+\bar\eta^{-2}(\wD\bar\omega)^2-2,\\ 
& G_{\bar\omega}(\tau)= \wD\left(\bar\eta^{-2} \wD\bar\omega\right),
\end{align}
\begin{align}
& G'_{\bar\sigma}(\tau)=\widehat \Delta \tilde\sigma + \bar\eta^{-2}(2 \wD \tilde\omega .
\wD\bar\omega - 2\tilde\sigma (\wD \bar \omega)^{2}),\\ 
& G'_{\bar{\omega}}(\tau)=\wD(\bar\eta^{-2}(\wD\tilde\omega -2\tilde\sigma \wD\bar\omega)),
\end{align}
and
\begin{align}
& F(\tau)=(\wD\tilde\sigma+\tilde\omega\bar\eta^{-2}\wD\bar\omega)^2+(\wD(\tilde\omega \bar\eta^{-1}-
\bar\eta^{-1}\tilde\sigma \wD\bar\omega))^2+(\bar\eta^{-1}\tilde\sigma \wD\bar\omega-\tilde\omega \bar\eta^{-2}\wD\bar\eta)^2,\\
& H(\tau)= \wD(\tilde\sigma \wD\tilde\sigma+\tilde\omega \bar\eta^{-1}\wD(\tilde\omega \bar\eta^{-1})).
\end{align}
Now we can use the expressions for $G'_{\bar\sigma}$ and $G'_{\bar\omega}$ to obtain, after a simple integration by parts,
\be
\partial_{\tau}^{2}\olfm_{ab}=-2\int_{\bar\theta_{a}}^{\bar\theta_{b}}(\tilde\sigma G'_{\bar\sigma}+\tilde\omega G'_{\bar\omega}-
\frac{1}{2}\tilde\sigma^2 \fv \eexp^{-\bar\sigma})\sin\bar\theta d\bar\theta.
\ee
Using \eqref{Carter}, integrating by parts once again  and using the boundary conditions $\tilde{\sigma}(\bar\theta_{a})=
\tilde{\sigma}(\bar\theta_{b})=0$, $\tilde{\omega}(\bar\theta_{a})=\tilde{\omega}(\bar\theta_{b})=0$ to get rid of $H$, 
yields
\be
\partial^{2}_{\tau}\olfm_{ab}=2\int_{\bar\theta_{a}}^{\bar\theta_{b}}(F+2\tilde\sigma\tilde\omega G_{\bar\omega}-
\bar\eta^{-2}\tilde\omega^2 G_{\bar\sigma}+\frac{1}{2}\tilde\sigma^2 \fv \eexp^{-\bar\sigma})\sin\bar\theta d\bar\theta.
\ee
Evaluating at $\tau=0$ and using the Euler-Lagrange equations, we obtain
\be
\partial^{2}_{\tau}\olfm_{ab}\big|_{\tau=0}=2\int_{\bar\theta_{a}}^{\bar\theta_{b}} (F+\frac{1}{2}(\bar\eta_c^{-2}\tilde\omega^2 +
\tilde\sigma^2)\fv\eexp^{-\bar\sigma_{c}})\sin\bar\theta d\bar\theta,
\ee
which can be written in the form
\begin{align}
\partial_{\tau}^{2}\olfm_{ab}\big|_{\tau=0}=\label{LIN} 2\int_{\theta_{a}}^{\theta_{b}}
\bigg\{\ & \bigg(\tilde{\sigma}'+\bigg(\frac{{\omega_{c}}'}{\bar\eta_{c}}\bigg)\bigg[\frac{\tilde{\omega}}{\bar\eta_{c}}\bigg]\bigg)^{2}+\bigg(\bigg[\frac{\tilde{\omega}}{\bar\eta_{c}}\bigg]'-\bigg(\frac{\omega_{c}'}{\bar\eta_{c}}\bigg)\tilde{\sigma}\bigg)^{2} \\ 
\label{F1} & +\bigg(\frac{\tilde{\sigma}\omega_{c}'}{\bar\eta_{c}}-\frac{\tilde{\omega}\bar\eta_{c}'}{\bar\eta_{c}^{2}}\bigg)^{2}+
 \frac{\fv}{2}\bigg(\bigg[\frac{\tilde{\omega}}{\bar\eta_{c}}\bigg]^{2}+\tilde{\sigma}^{2}\bigg)\eexp^{-\bar\sigma_{c}}\ \bigg\}\sin\bar\theta d\bar\theta.
\end{align}
We proceed by taking advantage of this formula.

\vs
First we note that because $\bar\gamma_c$ is a critical point we have $\omega_c'/\bar\eta_c^2=k/\sin\bar{\theta}$ where $k$ is a constant. Write 
$\tilde{\tilde\omega}:=\tilde\omega/\bar\eta_c$ and disregard the first term in \eqref{F1}. We get
\begin{equation}
 \label{F2}
\partial^{2}_{\tau}\olfm_{ab}\big|_{\tau=0}\geq 
2\int_{\bar{\theta}_{a}}^{\bar{\theta}_{b}}\bigg\{ \bigg(\tilde{\sigma}'+
\bigg(\frac{k\bar\eta_c}{\sin\bar{\theta}}\bigg)\tilde{\tilde{\omega}}\bigg)^{2}
+\bigg(\tilde{\tilde{\omega}}'-\bigg(\frac{k\bar\eta_c}{\sin\bar{\theta}}\bigg)\tilde{\sigma}\bigg)^{2} 
+\frac{\fv}{2}(\tilde{\tilde{\omega}}^2+\tilde{\sigma}^{2})\eexp^{-\bar{\sigma}_{c}}\ \bigg\}\sin\bar\theta d\bar\theta.
\end{equation}
Let $s:=\min\big\{(\sin\bar\theta)/\bar\eta_c\big\}$ and assume
\be\label{sup1}
\int_\Omega\tilde\sigma'^2\sin\bar\theta d\bar\theta>\frac{4k^2}{s^2}\int_\Omega\tilde{\tilde\omega}^2\sin\bar\theta d\bar\theta
\ee
Then the first term in \eqref{F2} can be bounded as
\begin{align}
\bigg[\int_{\bar\theta_{a}}^{\bar\theta_{b}}\left(\tilde\sigma'+\bigg(\frac{k\bar\eta_c}{\sin\bar\theta}\bigg)\tilde{\tilde{\omega}}\right)^2 
& \sin\bar\theta d\bar\theta \bigg]^{1/2}\geq \\ 
& \geq \bigg[\int_{\bar\theta_{a}}^{\bar\theta_{b}}\tilde\sigma'^2\sin\bar\theta d\bar\theta\bigg]^{1/2}-
\bigg[\int_{\bar\theta_{a}}^{\bar\theta_{b}}\bigg(\frac{k^2\bar\eta_c^2}{\sin^2\bar\theta}\bigg)\tilde{\tilde\omega}^2\sin\bar\theta d\bar\theta\bigg]^{1/2}\\
& \geq\bigg[\int_{\bar\theta_{a}}^{\bar\theta_{b}}\tilde\sigma'^2\sin\bar\theta d\bar\theta\bigg]^{1/2}-
\frac{|k|}{s}\bigg[\int_{\bar\theta_{a}}^{\bar\theta_{b}}\tilde{\tilde\omega}^2\sin\bar\theta d\bar\theta\bigg]^{1/2}\\
\label{eq1} &\geq\bigg[\int_{\bar\theta_{a}}^{\bar\theta_{b}}\tilde\sigma'^2\sin\bar\theta d\bar\theta\bigg]^{1/2}-
\frac{1}{2}\bigg[\int_{\bar\theta_{a}}^{\bar\theta_{b}}\tilde\sigma'^2\sin\bar\theta d\bar\theta\bigg]^{1/2}\\
& =\frac{1}{2}\bigg[\int_{\bar\theta_{a}}^{\bar\theta_{b}}\tilde\sigma'^2\sin\bar\theta d\bar\theta\bigg]^{1/2}
\geq \min\{\frac{\sin^{1/2}\bar\theta}{2}\}\bigg[\int_{\bar\theta_{a}}^{\bar\theta_{b}}\tilde\sigma'^2 d\bar\theta\bigg]^{1/2},
\end{align}
where \eqref{eq1} has been obtained
using \eqref{sup1}. This bound together with the last term in \eqref{F2}
gives  us
\be
\label{boundsigma}
\partial^{2}_{\tau}\olfm_{ab}\big|_{\tau=0}\geq c_{1}\|\tilde\sigma\|^2_{1,2}
\ee
for some constant $c_{1}>0$.

Now assume that the opposite to \eqref{sup1} holds, namely
\be\label{sup2}
\int_\Omega\tilde\sigma'^2\sin\bar\theta d\bar\theta\leq\frac{4k^2}{s^2}\int_\Omega\tilde{\tilde\omega}^2\sin\bar\theta d\bar\theta.
\ee
Then from \eqref{F2} we have
\begin{align}\label{F3}
\partial_{\tau}^{2}\olfm_{ab}|_{\tau=0}&\geq
\int_{\Omega}\fv(\tilde{\tilde{\omega}}^2+\tilde{\sigma}^{2})\eexp^{-\bar\sigma_{c}}\sin\bar\theta d\bar\theta\\
&\geq\min\{\fv\eexp^{-\bar\sigma_{c}}\}\int_{\bar\theta_{a}}^{\bar\theta_{b}}(\tilde{\tilde{\omega}}^2+\tilde{\sigma}^{2})\sin\bar\theta d\bar\theta\\
&\geq \min\{\fv\eexp^{-\bar\sigma_{c}}\}\int_{\bar\theta_{a}}^{\bar\theta_{b}}(\frac{s^2}{4k^2}\tilde\sigma'^{2}+\tilde\sigma^2)\sin\bar\theta d\bar\theta \\ 
&\geq \min\{\fv\eexp^{-\bar\sigma_{c}}\}\min\{1, \frac{s^2}{4k^2}\}\int_{\bar\theta_{a}}^{\bar\theta_{b}}(\tilde\sigma'^{2}+\tilde\sigma^2)\sin\bar\theta d\bar\theta
\end{align}
which again gives us an inequality $\partial^{2}_{\tau}\olfm_{ab}|_{\tau=0}\geq c_{2}\|\tilde\sigma\|^2_{1,2}$ 
for some constant $c_{2}>0$. Thus in either case we have 
\be\label{boundsigma2}
\partial^{2}_{\tau}\olfm_{ab}\big|_{\tau=0}\geq c_{3}\|\tilde\sigma\|^2_{1,2}
\ee
 for some constant $c_{3}>0$.

Now we can interchange the roles of $\tilde\sigma$ and $\tilde{\tilde\omega}$
(observing the symmetry in (\ref{F2})) to find again
\be\label{boundomega}
\partial^{2}_{\tau}\olfm_{ab}\big|_{\tau=0}\geq c_{3}\|\tilde{\tilde{\omega}}\|^2_{1,2}.
\ee
Using that $\tilde{\tilde{\omega}}=\tilde\omega/\eta_c$ and by an argument similar to the previous one we deduce from (\ref{boundomega}) that
\be\label{boundtomega}
\partial^{2}_{\tau}\olfm_{ab}\big|_{\tau=0}\geq c_{4}\|\tilde{\omega}\|^2_{1,2}
\ee
for some constant $c_{4}>0$. Collecting \eqref{boundsigma2} and \eqref{boundtomega} we get 
\be\label{boundM2}
\partial^{2}_{\tau}\olfm_{ab}\big|_{\tau=0}\geq c_{5}\|(\tilde\sigma,\tilde{\omega})\|^2_{1,2}=c_{5}\|\bar\gamma-\bar\gamma_c\|^2_{1,2}
\ee
for some constant $c_{5}>0$.

Having treated the second term on the right hand side of (\ref{TAY}) we turn to the last one. 
We claim that there is a constant $c_{6}>0$ such that if $\|\bar\gamma-\bar{\gamma}_{c}\|_{1,2}\leq 1$ then 
\be\label{BOUND}
\partial^{3}_{\tau}\olfm_{ab}\big|_{\tau=\tau^{*}}\leq c_{6}\|(\tilde{\sigma},\tilde{\omega})\|_{1,2}^{3}.
\ee
Combined with (\ref{TAY}) this would show, as we want, that if $\|(\tilde{\sigma},\tilde{\omega})\|_{1,2}\leq \epsilon$  
for $\epsilon$ sufficiently small, then (\ref{PROPINEQ}) holds for some constant $c>0$. The bound (\ref{BOUND}) is indeed 
easily obtained. A direct computation gives
\be
\partial^{3}_{\tau}\olfm_{ab}=-2\int_{\bar\theta_{a}}^{\bar\theta_{b}} \bigg(\frac{6\tilde{\sigma}\tilde{\omega}'^{2}- 
12\tilde{\sigma}^{2}\tilde{\omega}'\bar\omega'+
4\tilde{\sigma}^{3}\bar\omega'^{2}}{\bar\eta^{2}}+\frac{\fv}{2}{\tilde\sigma}^{3}\eexp^{-\tilde\sigma}\bigg)\sin\bar\theta d\bar\theta
\ee
Bounds for each  term in this integral, 
compatible with (\ref{BOUND}), are obtained by using that $\|\tilde{\sigma}\|_{\infty}\leq \sqrt{\pi}\|\tilde{\sigma}\|_{1,2}\leq 
\sqrt{\pi} \|(\tilde{\sigma},
\tilde{\omega})\|_{1,2}$, and that if $\|\bar\gamma-\bar{\gamma}_{c}\|_{1,2}\leq 1$ then $\|\bar\sigma\|_{\infty}\leq c_{7}$ and 
$\|\bar\omega'\|_{2}\leq c_{8}$ for constants $c_{7}>0$ and $c_{8}>0$. For instance the first term is bounded as 
\be
\bigg|12\int_{\bar\theta_{a}}^{\bar\theta_{b}}\frac{\tilde{\sigma}\tilde{\omega}'^{2}}{\bar\eta^{2}}\sin\bar\theta d\bar\theta\bigg|\leq 
12 \sup\big\{\frac{1}{\sin^{3}\theta}\big\}\, e^{2c_{7}}\|\tilde{\sigma}\|_{\infty}\|\tilde{\omega}'\|^{2}_{2}\leq 
c_{9}\|(\tilde{\sigma},\tilde{\omega})\|_{1,2}^{3}
\ee
for some constant $c_{9}>0$. The other terms are bounded in the same way.
\end{proof}

\begin{proof}[\bf Proof of Proposition \ref{GLOBAMIN}.]

 It will be more convenient to work with the functional $\fm_{ab}^{*}(\gamma^{*})$ of the arguments $\gamma^{*}=(u,\omega)$ with $u=-\ln\eta$, given by
\be\label{FMS}
\fm_{ab}^{*}(\gamma^{*})=\int_{\bar\theta_{a}}^{\bar\theta_{b}} (u'^{2}+\omega'^{2}e^{2u} + \fv^{*} e^{u})\,\sin\bar\theta d\bar\theta
\ee
where
\be
\fv^{*}=\fv \sin^{2}\bar{\theta}.
\ee
This functional is equal to $\fm_{ab}(\gamma)$ plus a constant independent of the
arguments. (Use $u=-\ln \eta$ in (\ref{OLFM})).

If $\fm^{*}_{ab}$ is shown to satisfy the Palais-Smale (PS) condition (see below), then a simple application of Proposition 
\ref{LOCMIN}  and the mountain pass theorem, 
as explained in the {\it Corollary} on page 187 of \cite{MR766489}, 
shows that $\gamma_{e}^{*}=(\ln \eta_{e},\omega_{e})$ is the only critical point and that $\fm^{*}_{ab}(\gamma_{e}^{*})$ 
is the strict absolute minimum of $\fm^{*}_{ab}$.

We explain now how to verify the PS condition. Recall first that the PS condition holds iff
any sequence $\gamma^{*}_{i}$ for which 
$\fm^{*}_{ab}(\gamma^{*}_{i})$ is bounded and for which $\|\delta \fm^{*}_{ab} (\gamma^{*}_{i})\|\rightarrow 0$ has a (strongly) 
convergent subsequence. 
Here $\|\delta \fm^{*}_{ab} (\gamma^{*}_{i})\|$ is the norm of the differential of $\fm^{*}_{ab}$ at
$\gamma^{*}_{i}$. Recall that this norm is $\|\delta \fm_{ab}^{*}(\gamma^{*})\| = 
\sup\big\{|\delta_{X} \fm^{*}_{ab}(\gamma^{*})|: \|X\|_{1,2}=1\big\}$. Note from this definition
 that if $\|\delta \fm_{ab}^{*}(\gamma^{*}_{i})\|\rightarrow 0$, then for any sequence $X_{i}$
with $\|X_{i}\|_{1,2}\leq K$  we have 
\be\label{TOMEN}
|\delta_{X_{i}} \fm_{ab}^{*}(\gamma^{*}_{i})|\rightarrow 0. 
\ee
Now, for any tangent vector $X=(\tilde{u},\tilde{\omega})$ to a point $\gamma^{*}=(u,\omega)$ we compute
\be\label{FVAR}
\delta_{X}\fm_{ab}^{*} (\gamma^{*}) = \int_{\bar\theta_{a}}^{\bar\theta_{b}} (2\tilde{u}'u'+2 \tilde{u}\omega'^{2} e^{2u} + 
2\tilde{\omega}' \omega'  e^{2u}+ \tilde{u} \fv \sin^{2}\bar{\theta} e^{u})\, \sin\bar\theta d\bar\theta. 
\ee
This expression will be used below.

Let $\gamma^{*}_{i}$ be a sequence such that $\fm_{ab}^{*}(\gamma^{*}_{i})$ is uniformly bounded and such that 
$\|\delta \fm_{ab}^{*}(\gamma_{i}^{*})\|\rightarrow 0$. From (\ref{FMS}) we deduce that $\|u'_{i}\|_{2}$ is uniformly bounded
\footnote{Note that there are constants $0<c_{1}<c_{2}<\infty$ such that $c_{1}<\sin\theta<c_{2}$.} and from 
this and (\ref{REFBEL}) that $u_{i}$ is uniformly bounded and uniformly continuous. By
the theorem of Arzel\`a -Ascoli, 
$u_{i}$ has a $C^{0}$-convergent subsequence (that we still index by `$i$'). As $\|u_{i}\|_{1,2}$ is 
uniformly bounded we can assume that $u_{i}$ converges weakly in $H^{1,2}$ too. 
Then, from the $C^{0}$-boundedness of $u_{i}$ and again from (\ref{FMS}), 
we deduce in a similar fashion that $\omega_{i}$ has a subsequence converging in $C^{0}$ and weakly in $H^{1,2}$. 

Assume then without loss of generality that for the above sequence $\gamma_{i}^{*}$
we have $u_{i}\rightarrow u_{\infty}$ and $\omega_{i}\rightarrow \omega_{\infty}$ weakly in $H^{1,2}$ and 
strongly in $C^{0}$. Let $c>0$ be a constant such that $c<e^{2u_{i}}\sin\bar\theta$ for all $i$. Then,
\begin{align}
\nonumber & c\int_{\bar\theta_{a}}^{\bar\theta_{b}} (\omega'_{i}-\omega'_{\infty} )^{2}d\bar\theta\leq 
\int_{\bar\theta_{a}}^{\bar\theta_{b}} (\omega'_{i}-\omega_{\infty}')^{2}e^{2u_{i}}\,\sin\bar\theta d\bar\theta=\\
\label{Z} & = \bigg( \int_{\bar\theta_{a}}^{\bar\theta_{b}} \omega_{i}'(\omega'_{i}-\omega_{\infty}') e^{2u_{i}}\,
\sin\bar\theta d\bar\theta - 
\int_{\theta_{a}}^{\bar\theta_{b}} \omega_{\infty}'(\omega'_{i}-\omega_{\infty}') e^{2u_{i}}\,\sin\bar\theta 
d\bar\theta \bigg)\rightarrow 0
\end{align}
where the first integral in (\ref{Z}) is seen to go to zero by taking $V_{i}=(0,\tilde{\omega}_{i})$ with 
$\tilde{\omega}_{i}=\omega_{i}-\omega_{\infty}$ in (\ref{TOMEN}), 
while the second integral in (\ref{Z}) tends to zero because $\omega_{i}\rightarrow \omega_{\infty}$ 
weakly in $H^{1,2}$ and $u_{i}\rightarrow u_{\infty}$ strongly in $C^{0}$ and weakly in $H^{1,2}$. 
From (\ref{REFBEL}) and (\ref{Z}) we deduce that $\|\omega_{i}-\omega_{\infty}\|_{2}\rightarrow 0$, 
which together with (\ref{Z}) again shows that $\omega_{i}\rightarrow \omega_{\infty}$ in $H^{1,2}$. 

The convergence $u_{i}\rightarrow u_{\infty}$ in $H^{1,2}$ is shown in the same fashion. \end{proof}

\begin{proof}[\bf Proof of Proposition \ref{Limit}.] 
Inequality (\ref{eqprop5.3a}) follows from Propositions \ref{LOCMIN} and \ref{GLOBAMIN}, 
together with the relation \eqref{MM} between the functionals $\fm_{ab}$ and $\olfm_{ab}$, as they imply 
that extreme KdS data $(\sigma_e,\omega_e)$ are the unique global minimisers of $\fm_{ab}$ among functions $(\sigma,\omega)$ having the same
boundary conditions as $(\sigma_e,\omega_e)$ at $\theta_a,\theta_b$.

The proof of \eqref{eqlimit} is line by line identical to the proof when $\Lambda=0$ and which was obtained in \cite{2011CQGra..28j5014A}. 
We will only sketch the argument here and refer the reader to \cite{2011CQGra..28j5014A} for details. It is important to remark that the 
presence of the cosmological constant plays no important role in this step.

Divide the interval $[0,\pi]$ in three regions, $\Omega_I=\{\sin\theta\leq e^{(\ln t)^2}\}$, $\Omega_{II}=\{e^{(\ln t)^2}\leq\sin\theta\leq t\}$ 
and $\Omega_{III}=\{t\leq\sin\theta\}$. Note that when $t$ goes to zero, the regions $\Omega_I$ and $\Omega_{II}$ shrink toward the poles, 
while $\Omega_{III}$ extends to cover the whole interval $[0,\pi]$. Then a specific partition function $f(\theta)$ (see eqs. (70)-(71) in 
\cite{2011CQGra..28j5014A})
is used to interpolate between extreme KdS horizon data 
in region $\Omega_I$ and general data in region $\Omega_{III}$. Define the auxiliary interpolating data $\gamma(t)=(\sigma(t),\omega(t))$ as
\be
\gamma(t)=f_t(\sin\theta)\gamma+(1-f_t(\sin\theta))\gamma_e,
\ee
then, as mentioned before, combining Propositions \ref{LOCMIN} and \ref{GLOBAMIN} on the region $[\theta_a,\theta_b]:=\Omega_{II}\cup\Omega_{III}$ for functions $\gamma(t):\Gamma_{ab}\to \mathbb R^2$ 
we find
\be\label{bound4}
\fm_{ab}(\gamma(t))\geq\fm_{ab}(\gamma_e).
\ee
Moreover, as $\gamma(t)|_{\Omega_I}=\gamma_e|_{\Omega_I}$, we can extend \eqref{bound4} to $[0,\pi]$ (recall that 
$[0,\pi]=\Omega_I\cup[\theta_a,\theta_b]$) to obtain
\be\label{boundep}
\fm(\gamma(t))\geq\fm(\gamma_e).
\ee
The final step is to show that as $t$ goes to zero, the mass functional for the auxiliary data converges to the
mass functional for the original general data, that is
\be\label{boundm}
\lim_{t\to0}\fm(\gamma(t))=\fm(\gamma).
\ee
This is done in an identical manner as in \cite{2011CQGra..28j5014A} (with $\Lambda$ being irrelevant here),
by using that $\omega=\omega_e+\mathcal O(\sin^2\theta)$ near the poles and that 
$\fm(\gamma)$ and $\fm(\gamma_e)$ are
well defined.

Inequalities \eqref{boundep} and \eqref{boundm} give \eqref{eqlimit}.

Moreover, using the explicit value
\be
\eexp^{\ \dfrac{\fm(\sigma_e,\omega_e,\hat{A},\hat{a})-\hat{\beta}}{8\hat\kappa}}=\frac{\hat A}{4\pi}
\ee
we find 
\be
\eexp^{\ \dfrac{\fm(\sigma,\omega,\hat{A},\hat{a})-\hat{\beta}}{8\hat\kappa}}\geq \frac{\hat A}{4\pi}
\ee
which is inequality \eqref{ineq4}.

\end{proof}
\section{Possible generalisations}
\label{sec:disc}

We conclude  discussing possible extensions of our main result to the case
with electromagnetic field and to the case $\Lambda < 0$.
In the former case we conjecture an inequality which, in addition to $A$, $J$ and $\Lambda$, 
 contains electric and magnetic charges  $Q_E$ and $Q_M$ in the combination  $Q^2=Q_E^2+Q_M^2$.
Such an extension is natural from the fact that all special cases are proven, 
in particular we recall \cite{Clement:2012vb} the bound  $A^2 \ge 16\pi^2 ( 4J^2 + Q^4)$ in the case $\Lambda
=0$. Moreover, extreme Kerr-Newman-deSitter saturates (\ref{J<}) and (\ref{Q<}).

\begin{Conjecture} 
Under the assumptions of Theorem \ref{main} but under the 
presence of an electromagnetic field with charges $Q_E$, $Q_M$ with 
$Q^2=Q_E^2+Q_M^2$ and for any $\Lambda > 0$ we have
\begin{equation}
\label{J<} 
J^2 \le  \frac{A^2}{64\pi^2} \left[
\left( 1 - \frac{\Lambda A}{4\pi} \right) \left( 1 - \frac{\Lambda A}{12\pi} \right)
- \frac{2\Lambda Q^2}{3} \right] - \frac{Q^4}{4}  
\end{equation}
or equivalently,
\begin{equation}
\label{Q<}
\left(Q^2 + \frac{\Lambda A^2}{48\pi^2} - 
\sqrt{\frac{A^2}{16\pi^2} \left(1 - \frac{\Lambda A}{6\pi} \right)^2 - 4 J^2 } \right)
\left(Q^2 + \frac{\Lambda A^2}{48\pi^2} + 
\sqrt{\frac{A^2}{16\pi^2} \left(1 - \frac{\Lambda A}{6\pi} \right)^2 - 4 J^2 } \right)
\le 0 
\end{equation}
Moreover, (\ref{J<}) and (\ref{Q<})
are saturated precisely for
extreme Kerr-Newman-deSitter configurations.

\end{Conjecture}

As to the calculations leading to (\ref{J<}) and
(\ref{Q<})  we made use of Equ. (44) of Caldarelli et al. \cite{Caldarelli:1999xj}, 
 where the temperature $T$ of a Kerr-Newman-anti-deSitter black hole is given in terms 
of $l^2 = -3/\Lambda$, the mass $M$, 
the entropy $S = A/4$, $Q$ and $J$. This calculation is insensitive to the
sign of $\Lambda$, and the requirement that $T \ge 0$ gives directly (\ref{J<}),
while (\ref{Q<}) is obtained via simple algebraic manipulations.

\vs

 We finally comment on the prospects of proving the area inequalities
 (\ref{A1}),  (\ref{J<}) and (\ref{Q<}) 
for the  case  $\Lambda < 0$ along the lines described above.
We first remark that extreme Kerr-anti-deSitter saturates (\ref{A1}) which should be
clear from the discussion of Sect. 3, and extreme Kerr-Newman-anti-deSitter saturates 
(\ref{J<}) and (\ref{Q<}).  
Next, the first part of our proof of (\ref{A1}), namely the 
lower bound for $A$ in terms on ${\cal M}$ as given in (\ref{ineq1})
carries over to $\Lambda < 0$ straightforwardly. 
However, attempts of obtaining a lower bound for ${\cal M}$ analogously to 
(\ref{ineq2}) seem to be in vain. The reason is that one can easily construct examples with sufficiently 
small $\sigma$, (negative with large modulus), and suitably adjusted $\omega$ 
for which the last term in (\ref{defM}), 
which is now negative, dominates the first two positive terms. 
In fact these examples strongly suggest that ${\cal M}$ is even unbounded from
below  unless the data are restricted appropriately.
Therefore, while it is still possible that (\ref{A1}), (\ref{J<}) and (\ref{Q<}) 
hold for $\Lambda < 0$ as well, our strategy which was successful for $\Lambda > 0$ is unlikely to carry over. 
     
\bigskip

{\bf Acknowledgements.} We acknowledge helpful discussions with
Lars Andersson, Piotr Bizo\'n, Piotr Chru\'sciel, Sergio Dain, Jose Luis
Jaramillo, Marc Mars and Luc Nguyen.

M.E.G.C. is supported by CONICET (Argentina). W.S. was funded by the Austrian
Science Fund (FWF): P23337-N16 and by the Albert Einstein Institute
(Potsdam).

\bibliographystyle{plain}

\bibliography{Master}

\end{document}